\documentclass[12pt]{article}

\usepackage{scicite}

\usepackage{times}

\usepackage{amsmath}
\usepackage{amsfonts}
\usepackage{amssymb}
\usepackage{graphicx}
\usepackage{comment}

\newenvironment{sciabstract}{%
\begin{quote} \bf}
{\end{quote}}

\title{Ultrafast Dynamical Lifshitz Transition}

\author
{Samuel Beaulieu$^{1\dagger}$*, Shuo Dong$^{1}$, Nicolas Tancogne-Dejean$^{2}$*, Maciej Dendzik$^{1,3}$,\\
Tommaso Pincelli$^{1}$, Julian Maklar$^{1}$, R. Patrick Xian$^{1}$, Michael A. Sentef$^{2}$,\\
Martin Wolf$^{1}$, Angel Rubio$^{2,4}$, Laurenz Rettig$^{1}$, and Ralph Ernstorfer$^{1}$*\\
\\
\normalsize{$^{1}$Fritz Haber Institute of the Max Planck Society,}\\
\normalsize{Faradayweg 4-6, 14195 Berlin, Germany}\\
\normalsize{$^{2}$Max Planck Institute for the Structure and Dynamics of Matter,}\\
\normalsize{Luruper Chaussee 149, 22761 Hamburg, Germany}\\
\normalsize{$^{3}$Department of Applied Physics, KTH Royal Institute of Technology,}\\
\normalsize{Institute of Technology, Electrum 229, SE-16440, Stockholm, Kista, Sweden}\\
\normalsize{$^{4}$Center for Computational Quantum Physics (CCQ),}\\
\normalsize{The Flatiron Institute, 162 Fifth Avenue, New York NY 10010}\\
\\
\normalsize{$^\dagger$Current affiliation:}\\
\normalsize{Université de Bordeaux - CNRS - CEA, CELIA, UMR5107, F33405, Talence, France}
\\
\normalsize{$^\ast$To whom correspondence should be addressed:}
\\
\normalsize{beaulieu@fhi-berlin.mpg.de}\\
\normalsize{nicolas.tancogne-dejean@mpsd.mpg.de}\\
\normalsize{ernstorfer@fhi-berlin.mpg.de}
}

\date{\today}

\begin{document} 


\baselineskip24pt


\maketitle 

\begin{sciabstract}
Fermi surface is at the heart of our understanding of metals and strongly correlated many-body systems. An abrupt change in the Fermi surface topology, also called Lifshitz transition, can lead to the emergence of fascinating phenomena like colossal magnetoresistance and superconductivity. While Lifshitz transitions have been demonstrated for a broad range of materials by equilibrium tuning of macroscopic parameters such as strain, doping, pressure and temperature, a non-equilibrium dynamical route toward ultrafast modification of the Fermi surface topology has not been experimentally demonstrated. Combining time-resolved multidimensional photoemission spectroscopy with state-of-the-art TDDFT+$U$ simulations, we introduce a novel scheme for driving an ultrafast Lifshitz transition in the correlated type-II Weyl semimetal T$\mathrm{_{d}}$-MoTe$_{2}$. We demonstrate that this non-equilibrium topological electronic transition finds its microscopic origin in the dynamical modification of the effective electronic correlations. These results shed light on a novel ultrafast scheme for controlling the Fermi surface topology in correlated quantum materials.
\end{sciabstract}

\section*{Introduction}

The free quantum electron gas approximation within the Drude-Sommerfeld model of metallic solids leads to purely parabolic band dispersion and spherical Fermi surfaces. However, electron-lattice and electron-electron interactions break the quadratic energy-momentum relationship, leading to complex electronic band dispersion's and Fermi surface shape and topology. The Fermi surface, which separates unoccupied from occupied electronic states, is of paramount importance to understand a broad range of phenomena in metallic and semimetallic systems. To cite pioneering work of Kaganov and Lifshitz: ``the Fermi surface is the stage on which the drama of the life of the electron is played out''\cite{Kaganov79}.

Equilibrium tuning of static macroscopic parameters such as temperature\cite{wu15,zhang17,chen20}, pressure\cite{Xiang15,kang15}, strain\cite{sunko19,burganov16}, external magnetic fields \cite{Ptok17} or doping\cite{Liu10,Shi17}, have all been demonstrated to be capable of modifying electronic band structures, which is, in some prominent cases, accompanied by an abrupt change in the Fermi surface topology, a phenomenon known as \textit{Lifshitz transition}\cite{Lifshitz60}. A Lifshitz transition is often concurrent with strong modifications of material properties since low-energy excitations, relevant to many electronic, magnetic, and optical properties, occur nearby the Fermi surface boundary. For example, such electronic topological transition often leads to a sudden change of the transport properties\cite{wang15,Chi17}. Moreover, it was demonstrated that Lifshitz transitions coincide with the onset of superconductivity in electron-doped iron arsenic superconductors\cite{Liu10}. While inducing Lifshitz transitions using equilibrium tuning methods is now well-established and understood, a protocol for ultrafast Fermi surface topology engineering would be highly desirable since it could allow controlling materials properties on unprecedented timescales. However, until now, this goal remains elusive. 

Here, we experimentally and theoretically demonstrate a route to induce ultrafast Lifshitz transitions in correlated materials, based on transient band structure modifications upon the interaction with ultrashort laser pulses. We apply this scheme to the topological type-II Weyl semimetal T$_{d}$-MoTe$_{2}$, a material that is known to be in the vicinity of a Coulomb-interaction-induced Lifshitz transition\cite{Xu18}. Experimentally, we directly map the dynamical evolution of its band structure and its Fermi surface on ultrafast timescales using state-of-the-art time-resolved multidimensional photoemission spectroscopy\cite{kutnyakhov20}. Theoretically, we investigate the microscopic origin of the non-equilibrium Lifshitz transition using extensive time-dependent self-consistent Hubbard $U$ calculations (TDDFT+$U$)\cite{Tancogne-Dejean17,Tancogne-Dejean18,Topp18}. In striking contrast to statically induced Lifshitz transitions, we here show a disruptive out-of-equilibrium strategy to control Fermi surface topology on unprecedented timescales.  

\begin{figure}
\centering\includegraphics[width=0.8\textwidth]{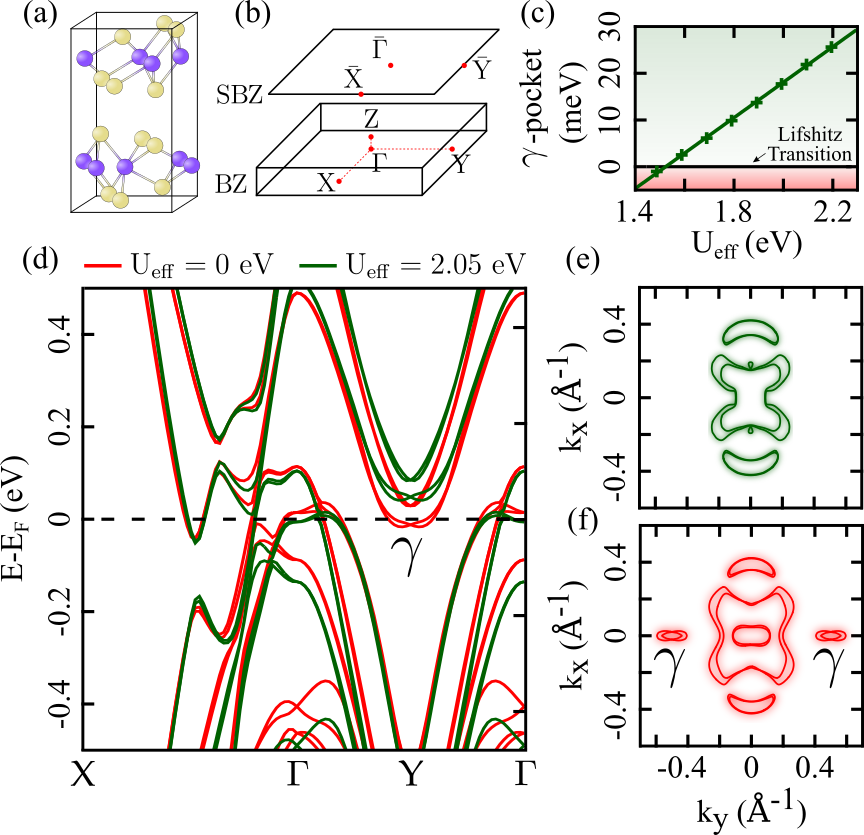}
\caption{\textbf{Coulomb-interaction-induced Lifshitz transition in topological Weyl semimetal $\mathrm{\mathbf{T_d}}$\textbf{-}$\mathrm{\mathbf{MoTe_2}}$}. (a) A schematic of the crystal structure of the low-temperature orthorhombic $\mathrm{T_d}$-phase of $\mathrm{MoTe_2}$ and (b) its associated Brillouin zone (BZ) and surface Brillouin zone (SBZ). (c) The equilibrium phase diagram revealing the effect of the Hubbard $U_{\mathrm{eff}}=U-J$ ($J$ being the Hund's exchange) on the position of the $\gamma$-pocket relative to the Fermi energy. The dots are data points obtained from DFT+$U$ simulations whereas the thick line is a linear fit to the data, serving as a guide to the eye. (d) The electronic band structure of $\mathrm{T_d}$-$\mathrm{MoTe_2}$ for the equilibrium self-consistently calculated value of $\mathrm{U_{eff}}$ = 2.05 eV (in green) and for the reduced value of $U_{\mathrm{eff}}$ = 0 eV (below the adiabatic Lifshitz transition), and their respective Fermi surface cuts taken at $k_z$ = 0, (e)-(f). In (f), one can clearly see the hallmark of the Coulomb-induced Lifshitz transition--the appearance of $\gamma$-electron pockets at the Fermi surface.}
\label{Fig1}
\end{figure}

$\mathrm{T_d}$-$\mathrm{MoTe_2}$ belongs to the transition metal dichalcogenide (TMDC) family. TMDCs have recently attracted tremendous interest, mostly because of the unique combination of strong spin-orbit splitting, locally broken inversion symmetry, and enhanced electronic and mechanical properties, making them very interesting for fundamental studies as well as for applications in electronics, spintronics, optoelectronics, and twistronics. In particular, when cooled below $\sim$ 250 K, $\mathrm{MoTe_2}$ is known to undergo a structural phase transition from the monoclinic and topologically trivial semimetallic (1T'-) phase to the orthorhombic type-II Weyl semimetallic ($\mathrm{T_d}$-) phase (Fig.~\ref{Fig1}(a)-(b)), characterized by tilted Weyl cones originating from a protected crossing between the valence and conduction bands in reciprocal space\cite{soluyanov15}. Weyl points act as a topological charge, \textit{i.e.} either a source or a sink of Berry curvature, leading to many fascinating physical phenomena, including the emergence of Fermi arcs \cite{wan11,xu15} as well as various transport anomalies \cite{hosur13}. Elucidating the detailed electronic structure of this prominent 2D Weyl semimetal has been subject of a lot of experimental and theoretical efforts. Recently, combined static soft X-ray angle-resolved photoemission spectroscopy (ARPES) and density functional theory (DFT)+$U$ studies reported that the inclusion of corrected on-site Coulomb interaction (Hubbard $U$) is essential to reproduce the experimentally measured electronic band structure and Fermi surface of $\mathrm{T_d}$-$\mathrm{MoTe_2}$\cite{Xu18,Aryal19}. 

\section*{Results and Discussion}

Fig.~\ref{Fig1}(d) shows the calculated band structures for different values of the effective Hubbard $\mathrm{U_{eff}}$: in green for the self-consistent value ($\mathrm{U_{eff}}$ = 2.05 eV) obtained using our first-principle method\cite{Tancogne-Dejean17}, and in red, for a reduced $\mathrm{U_{eff}}$ = 0 eV. While the modification of the effective on-site Coulomb interaction leaves the band dispersions along $\mathrm{\Gamma}$-X mostly unchanged, it leads to a significant energy shift of two slightly spin-orbit split electron pockets (separated by 36.5 meV), located around the Y high symmetry points. Fig.~\ref{Fig1}(c) shows the energy position of the lower-lying $\gamma$-pocket as a function of the effective Hubbard $U\mathrm{_{eff}}$. Around $U\mathrm{_{eff}}$ $\sim$ 1.5 eV, the $\gamma$-pocket is crossing the Fermi level, leading to a Lifshitz transition, as evidenced by the Fermi surface cuts depicted on Fig.~\ref{Fig1} (e)-(f), in good agreement with the predictions of Xu \textit{et al.}\cite{Xu18}. Because $\mathrm{T_d}$-$\mathrm{MoTe_2}$ is in the vicinity of a Coulomb-induced Lifshitz transition, it is an interesting candidate for investigating the possibility to control the Fermi surface topology on ultrafast timescales.

\begin{figure}
\centering\includegraphics[width=0.8\textwidth]{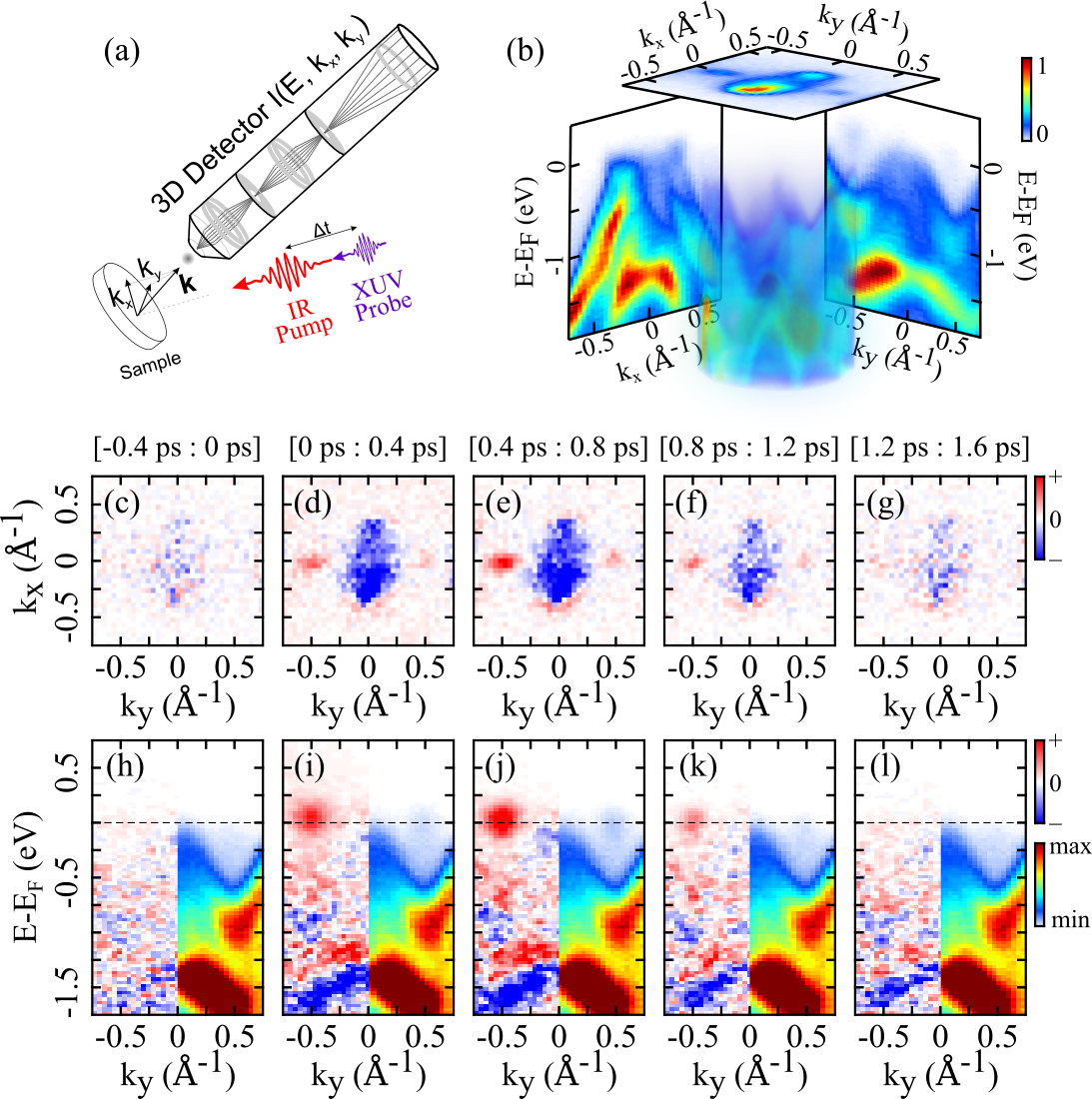}
\caption{\textbf{Ultrafast dynamical Lifshitz transition probed by time-resolved multidimensional photoemission spectroscopy.} (a) A schematic of the experimental setup, featuring IR-pump/XUV-probe pulses and a time-of-flight momentum microscope detector allowing for parallel measurement of the band structure of the crystalline solid, as a function of pump-probe time delay ($\mathrm{E_B}$, $\mathrm{k_x}$, $\mathrm{k_y}$, t). (b) An example of the experimental 3D volumetric photoemission data, as well as cuts along different high-symmetry directions, and cut at the Fermi energy, integrated for all positive time delays. (c)-(g) Differential (unpumped signal subtracted) 2D Fermi surfaces ($\mathrm{k_x}$, $\mathrm{k_y}$) as a function of time delay between the IR pump and the XUV probe (integrated over 400 fs intervals). (h)-(l) Corresponding raw (right, $\mathrm{k_y>0}$) and differential (left, $\mathrm{k_y<0}$) energy-resolved cuts along Y-$\mathrm{\Gamma}$-Y ($\mathrm{k_x=0}$).}
\label{Fig2}
\end{figure}

We used time-resolved multidimensional photoemission spectroscopy to directly probe the non-equilibrium electronic structure of $\mathrm{T_d}$-$\mathrm{MoTe_2}$. Our experimental setup includes a monochromatized high-order harmonic generation (HHG)-based XUV source, at 500~kHz repetition rate and centered around 21.7~eV (bandwidth: 110~meV FWHM, pulse duration: $\sim$20~fs FWHM)\cite{Puppin19}, spanning the full extent of the surface Brillouin zone in parallel momentum. Both IR-pump (1030 nm, 140 fs FWHM, 6.7x10$\mathrm{^9}$ $\mathrm{W/cm^2}$) and XUV-probe are $p$-polarized and focused onto the $\mathrm{T_d}$-$\mathrm{MoTe_2}$ sample, handled and cooled to 30 K by a 6-axis cryogenic manipulator. The photoemitted electrons are collected by a time-of-flight momentum microscope\cite{Medjanik17}, a multidimensional detection scheme to obtain the four-dimensional (E$_{B}$,k$_{x}$,k$_{y}$,t) non-equilibrium electronic structure (see Fig.~\ref{Fig2}(a)-(b))\cite{kutnyakhov20}.  We call this technique time-resolved multidimensional photoemission spectroscopy since we directly measure four-dimensional photoemission intensity I($\tau$, $E_B$, $k_x$, $k_y$), instead of more standard three-dimensional photoemission intensity I($\tau$, $E_B$, $k_{||}$), when using hemispherical analyzer. More information about the experimental setup can be found in Methods, and in \cite{Maklar20}.

We tracked the evolution of the Fermi surface as a function of pump-probe delay (Fig.~\ref{Fig2} (c)-(g)). After the interaction with the femtosecond IR pump pulse, the appearance of the $\gamma$-pockets is visible (Fig.~\ref{Fig2}(d)) on the Fermi surface. This is the hallmark of the dynamical Lifshitz transition. At longer delays, the signal in the $\gamma$-pockets at the Fermi energy continues to increase (Fig.~\ref{Fig2}(e)), followed by a gradual decays (Fig.~\ref{Fig2}(f)), and eventually vanish after $<$ 1 ps (Fig.~\ref{Fig2}(g)), indicating the recovery of the equilibrium Fermi surface topology. 

What is the origin of the experimentally observed transient change of the spectral weight on the Fermi surface? At first glance, few different scenarios could lead to such observation: (i) a simple light-induced population of the originally unoccupied $\gamma$-pocket concomitant with the finite energy resolution of the experimental setup, (ii) a Floquet-type state from periodically-driven lower-lying bands, (iii) a light-induced structural phase transition leading to the modification of the electronic structure, and finally, (iv) renormalization of the electronic band structure due to dynamical changes in the effective electronic correlations (transient changes of Hubbard $U$). In the following, we provide clear evidence that only the last scenario is in complete agreement with both our experimental observations and our first-principle simulations.

\begin{figure}
\centering\includegraphics[width=0.8\textwidth]{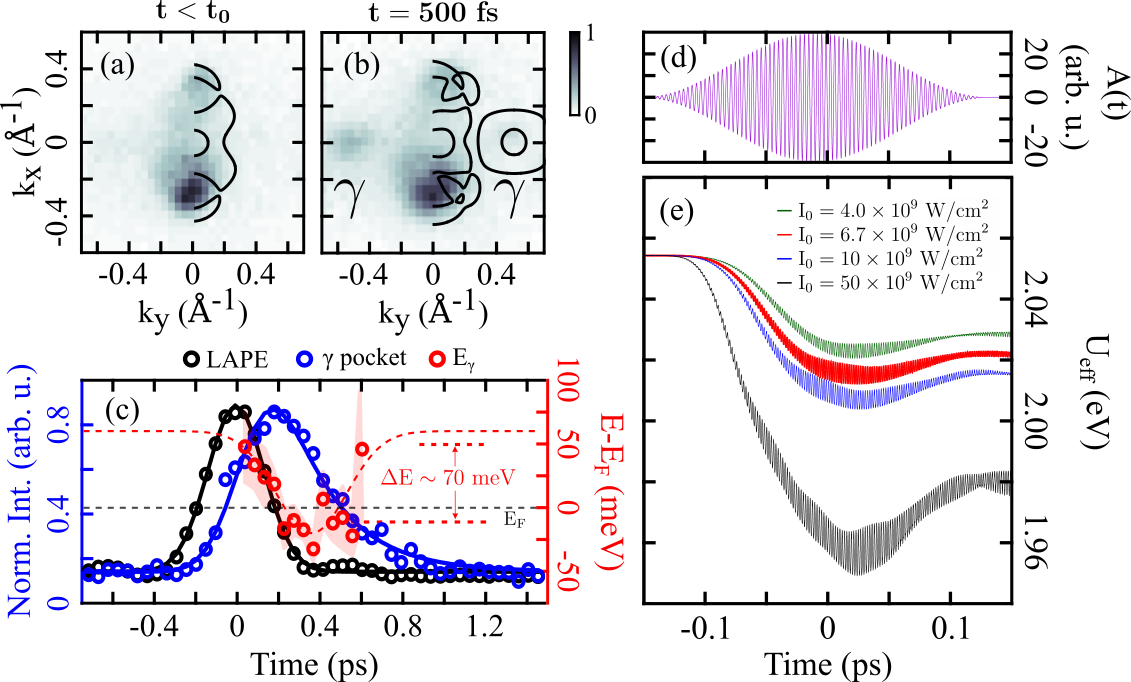}
\caption{\textbf{Ultrafast modification of Hubbard $U$ at the origin of the dynamical Lifshitz transition.} 
(a)-(b) Experimentally measured Fermi surface for $\mathrm{t < t_0}$ and t = 500 fs, and Fermi surface before and after the pump pulse, obtained by TDDFT+$U$ simulations using the same intensity as in the experiment ($\mathrm{I_0}$ = 6.7x$\mathrm{10^{9}}$ $\mathrm{W/cm^2}$). In both cases (theory and experiment), the $\gamma$-electron pocket close to the Y point, \textit{i.e.} the hallmark of the Lifshitz transition, is clearly showing up after the interaction with the pump pulse. (c) Normalized laser-assisted photoemission signal (LAPE) (black), normalized $\gamma$-pocket excited-state signal (blue), and energy position of the bottom of the $\gamma$-pocket (red), as a function of pump-probe delay. The shallow red curve represents the 95$\%$ confidence intervals of the $\gamma$-pocket position extracted from EDCs fitting. The dashed red curve serves as a guide to the eye. (d) The vector potential of the laser pulse used in the self-consistent TDDFT+$U$ simulations, with the same wavelength and duration as used in the experiment. (e) Dynamical evolution of $\mathrm{U_{eff}}$ for different laser fluences.}
\label{Fig3}
\end{figure}

The multidimensional nature of the experimental photoemission data allows us to gain an additional perspective on the laser-induced dynamics by looking at the energy-resolved signal along Y-$\mathrm{\Gamma}$-Y (see SM for time-dependent cuts along X-$\mathrm{\Gamma}$-X). These momentum-energy snapshots reveal the light-induced population dynamics within the $\gamma$-pockets accompanied, more importantly, by their time-dependent energy downshift. Indeed, by fitting the time-resolved energy distribution curves (EDCs) at the $Y$ point (see SM), we have extracted the position of the bottom of the $\gamma$-pocket, which is shown to transiently and reversibly downshift in energy by $\sim$ 70 meV (Fig.\ref{Fig3}(c)). This dynamical energy downshift of the $\gamma$-pockets results in a transient crossing of the ground state Fermi energy by 17 $\pm$ 7 meV, and causes a disruption of the Fermi surface topology. This effect can neither be explained by a transient excited-state population in a rigid band structure picture nor by the finite energy resolution of the experimental setup. This unambiguously rules out scenario (i). Moreover, scenario (ii), involving Floquet-type states, can also be safely ruled out since the measured renormalization of the $\gamma$-pocket is significantly delayed with respect to the pump pulse (see Fig.\ref{Fig3}(c)). 

It hence remains scenario (iii) and (iv) as plausible origins of the observed ultrafast Lifshitz transition. As shown in the SI, $\gamma$-pockets lying slightly below the Fermi energy is also a feature of the 1T' phase of $\mathrm{MoTe_2}$. A photoinduced structural phase transition from the $\mathrm{T_d}$ to the 1T' phase would, therefore, lead to the $\gamma$-pockets crossing the Fermi level upon photoexcitation, on the timescale of these structural changes. Zhang \textit{et al.} has recently studied this ultrafast lattice symmetry switching ($\mathrm{T_d}$ to 1T') using optical techniques\cite{zhang19}, by looking at the fluence dependence of coherent phonon modes and of the intensity loss of second-harmonic generation. When pumping above a critical fluence of $>$2 mJ/$\mathrm{cm^2}$, they showed that the structural phase transition occurs on the timescale of $\sim$700 fs and recovers to the $\mathrm{T_d}$ phase within hundreds of picoseconds. The absorbed pump fluence that we used ($\sim$ 0.6 mJ/$\mathrm{cm^2}$) is significantly lower than the structural phase transition critical fluence, and we observe a sub-400 fs modification of the Fermi surface topology with a sub-1.5 ps recovery timescale, which does not seem to be compatible with a photoinduced structural phase transition, as observed in Zhang \textit{et al.} \cite{zhang19}. We did not observed any signature of coherent phonons in our experimental data. Moreover, due to the longer wavelength of our pump pulse (1030 nm vs 800 nm), the penetration depth ($\delta$=410 nm) is more than 4 times larger than at 800 nm ($\delta$=100 nm). This means that the pulse energy is deposited on a much larger volume, leading a much smaller effective temperature increase. Indeed, using the equation $\mathrm{S\delta\rho/M \int_{T_0}^{T_0+\Delta T}C_p(T) dT = FS}$, where $\mathrm{S}$, $\mathrm{\rho}$, $\mathrm{M}$, $\mathrm{C_p}$ and $\mathrm{F}$ are respectively the excitation area, the mass density, the molar mass, the heat capacity, and the absorbed fluence, we can estimate the lattice temperature rise $\mathrm{\Delta T}$ after the interaction with the pump~\cite{zhang19}. Using the experimental values of $\mathrm{\rho}$, $\mathrm{M}$ and $\mathrm{C_p}$~\cite{kiwia1975low}, we obtain that our laser excitation rises the lattice temperature from $\mathrm{T_0}$ = 30~K to $\mathrm{T_0+\Delta T}$ = 71~K, which is well below the critical structural phase transition temperature of $\sim$ 250~K. Therefore, the photoinduced structural phase transition (scenario (iii)) can be safely excluded.

The remaining scenario which fits with all experimental observations is that the ultrafast Lifshitz transition is of electronic nature and originates from the dynamical modification of $U$. To investigate this scheme in greater details, we have performed first-principles self-consistent TDDFT+$U$ calculations \cite{Tancogne-Dejean17}. This method captures \textit{both} the laser-induced dynamical populations (non-thermal distribution of states) and the time-dependence of the effective electronic correlations (dynamical Hubbard $U$). For these calculations, we assume a frozen lattice.

 The calculated Fermi surfaces before and after the interaction with the pump pulse, shown in Fig.~\ref{Fig3}(a)-(b), clearly show that the TDDFT+$U$ simulations predict the ultrafast dynamical Lifshitz transition, in agreement with the experimental observations. Moreover, as predicted for the strongly correlated charge-transfer insulator NiO\cite{Tancogne-Dejean18} and a pyrochlore iridate\cite{Topp18}, the Hubbard $U$ is found to decrease upon photoexcitation for $\mathrm{T_d}$-$\mathrm{MoTe_2}$ (Fig.~\ref{Fig3}(d)-(e)). This can be understood in terms of dynamical enhancement of the electronic screening due to the delocalized nature of the pump-induced excited electrons\cite{Tancogne-Dejean18}. In the simulations, the modification of the Hubbard $U$ seems to reach a plateau after the end of the laser pulse (Fig.~\ref{Fig3} (e)), while in the experiment, the energy shift of the pockets is delayed compared to the pump laser pulse, timed by the LAPE signal (Fig.~\ref{Fig3} (c)). This may be understood as the effect of other scattering processes, such as electron-phonon coupling, unaccounted for in the simulation, that will prevent the laser-driven quenching of $U$ from persisting indefinitely. In other words, the simulations intrisically cannot capture the energy upshift of the pocket at longer pump-probe delays. Moreover, we note that the good agreement (for small pump-probe delays) between theory, which assumes a frozen lattice, and experiment, in describing the lowering of the $\gamma$ pocket is yet another indication that coherent phonons are not playing an important role observed dynamics.

Another interesting observation is that the calculated dynamical reduction of the $\mathrm{U_{eff}}$ (Fig.~\ref{Fig3} (e)) is found to be much smaller than what is expected for reaching the Lifshitz transition from equilibrium calculations (Fig.~\ref{Fig1} (c)). To explain why such a small change in $\mathrm{U_{eff}}$ can induce a Lifshitz transition in the non-equilibrium case, whereas this would not be the case for an adiabatic (equilibrium) scenario, we have to remember that both dynamical populations and time-dependent Hubbard $U$ can strongly affect the electronic structure and the Fermi surface of $\mathrm{T_d}$-$\mathrm{MoTe_2}$. We, therefore, investigated their respective role in driving the non-equilibrium Lifshitz transition (see SM). From this detailed analysis, we found that modifications of \textit{both} dynamical populations and Hubbard $U$ are required to reach the ultrafast dynamical Lifshitz transition. Indeed, if we freeze the Hubbard $U$ in the simulations, the Lifshitz transition does not occur. We also found that using the adiabatic electronic states to define the Fermi surface after laser excitation does not lead to the Lifshitz transition. 

Our TDDFT+$U$ framework, which assumes a frozen lattice, cannot quantitatively reproduce the full dynamics from photoexcitation to thermalization, \textit{e.g.} because of the lack of electron-phonon coupling. However, qualitatively, the fact that this novel non-equilibrium route requires a significantly smaller modification of the Hubbard $U$ to reach the Lifshitz transition has strong implications. First, it reveals that non-adiabaticity plays a key role in the observed ultrafast Lifshitz transition. Moreover, our results establish that the synergy between dynamical populations and Hubbard $U$ can facilitate reaching topological electronic transitions, in cases where only changing $U$, using adiabatic techniques, would not allow it. 

In conclusion, using time-resolved multidimensional photoemission spectroscopy, we have demonstrated a fundamentally new non-equilibrium scheme that allows us to drive an ultrafast Lifshitz transition in the Weyl semimetal T$_{d}$-MoTe$_{2}$, upon transient and reversible modification of the band structure. Our first-principle simulations revealed the role of the dynamical modulation of the populations and, more importantly, of the electronic correlations, in reaching the topological electronic transition on ultrafast timescales. Our work thus demonstrates that dynamical correlations and non-adiabaticity are key ingredients to drive the non-equilibrium Lifshitz transition. This ultrafast topological transition thus finds its roots in different physical mechanisms than more conventional adiabatic Lifshitz transitions. Moreover, the ultrafast Lifshitz transition presented here is of electronic origin, allowing to switch between different Fermi surface topologies, and thus to switch material's (\textit{e.g.} transport) properties\cite{chen16}, on very fast timescales. Combining this scheme with the emerging field of twistronics\cite{yankowitz19} will allow having an unprecedented level of control on the electronic correlations, on timescales not accessible to established adiabatic methods.   

\section*{Materials and Methods}

\textbf{Time-resolved multidimensional photoemission spectroscopy}

The time-resolved multidimensional photoemission spectroscopy experiments were performed at the Fritz Haber Institute of the Max Planck Society. We used a home-built optical parametric chirped-pulse amplifier (OPCPA) delivering 15 W (800 nm, 30 fs) at 500 kHz repetition rate. The second harmonic of the OPCPA output (400 nm) is used to drive high-order harmonic generation (HHG) by tightly focusing (15 $\mu$m FWHM) p-polarized laser pulses onto a thin and dense Argon gas jet. The extremely nonlinear interaction between the laser pulses and the Argon atoms leads to the generation of a comb of odd harmonics of the driving laser, extending up to the 11th order. The copropagating driving laser and the harmonics are reflected onto a silicon wafer at Brewster's angle of the 400 nm to filter out the energy of the fundamental driving laser. Next, a single harmonic (7th order, 21.7 eV) is isolated by reflection on a focusing multilayer XUV mirror and propagation through a 400 nm thick Sn metallic filter. A photon flux of up to 2x10$^{11}$ photons/s at the sample position is obtained (110 meV FWHM)\cite{Puppin19}. 

As a pump beam, we used a fraction of the compressed 1030 nm pulses used to generate the white-light seed in the OPCPA. The combination of longer wavelength and longer pulse duration (compared with 800 nm, 30 fs, typically used) allows us to use higher pump fluence before reaching the threshold of pump-induced multiphoton photoemission, which introduces space-charge distortion in the time-resolved multidimensional photoemission measurement. Using high-enough fluence is essential to drive the non-equilibrium Lifshitz transition. The drawback is that the temporal resolution decreases to $\sim$ 140 fs FWHM. The pump beam was also linearly p-polarized (along the $\mathrm{\Gamma}$- X direction of the crystal) and was at an angle of incidence of 65$^\circ$ to the sample surface normal. 

The bulk $\mathrm{T_d}$-$\mathrm{MoTe_2}$ samples are cooled to 30 K on the 6-axis cryogenic manipulator (SPECS GmbH) and cleaved at a base pressure of 2x10$^{-11}$ mbar. The data are acquired using a \textit{so-called} time-of-flight momentum microscope (METIS100, SPECS GmbH), allowing to detect each photoelectron as a single event and as a function of pump-probe delay. The resulting 4D photoemission intensity data have the coordinates I($E_B$, $k_x$, $k_y$, $t$). This represents an improvement with respect to standard time- and angle-resolved photoemission spectroscopy, where 3D photoemission intensity \textit{i.e.} I($E_B$, $k_{||}$,$t$) is typically measured, and where $k_{||}$ is the parallel momentum of the electron in the crystal along one specific direction of the Brillouin zone. The measured $k_{||}$ is determined by the orientation of the crystal with respect to the hemispherical analyzer slit. 

Concerning the data post-processing, the combination of the high repetition rate of our beamline and the multidimensional data recording scheme\cite{kutnyakhov20} leads to typical datasets involving 10$^9$-10$^{11}$ detected events and a typical dataset size of few hundreds of gigabytes (GBs). We thus use a recently developed open-source workflow\cite{xian19} to efficiently convert these raw single-event-based datasets into binned calibrated data hypervolumes of the desired dimension, including axes calibration and detector artifact corrections. Binning of the single-event data to a specific hypervolume reduces the data to a manageable size of a few to tens of gigabytes.

\textbf{DFT+$U$ and TDDFT+$U$ calculations}

All the calculations presented in the article, as well as in the Supplementary Information (SI), were performed for bulk $\mathrm{T_d}$-$\mathrm{MoTe_2}$, unless stated otherwise. Calculations were performed using fully relativistic Hartwigsen-Goedecker-Hutter (HGH) norm-conserving pseudo-potentials, using the crystal structure taken from\cite{Qi16} (real-space spacing of $\Delta r=0.158 $  $\mathrm{\AA}$). Time-dependent calculations have been performed using a $6\times5\times3$ $\mathbf{k}$-point grid to sample the Brillouin zone, whereas ground-state calculations were performed using a $12\times10\times6$ $\mathbf{k}$-point grid. We found that this choice does not lead to a sizable change in the value of the computed Hubbard $U_{\mathrm{eff}}$.
The pump driving field is taken along the [100] crystallographic direction in all the calculations, which corresponds to the $\mathrm{\Gamma}$-X direction. We considered a laser pulse of 140 fs duration (FWHM), with a sine-square envelope for the vector potential. The carrier wavelength $\lambda$ is 1030\,nm. The time-dependent wavefunctions and Hubbard $U_{\mathrm{eff}}$ are computed by propagating generalized Kohn-Sham equations within real-time TDDFT+$U$, as provided by the Octopus package\cite{1912.07921,Andrade15,castro06}. We employed the local density approximation for describing the local DFT part, and we computed the effective $U_{\mathrm{eff}}=U-J$ for Mo $d$ orbitals, using localized atomic orbitals from the corresponding pseudopotentials\cite{Tancogne-Dejean17}.

We used the real-time TDDFT plus $U$ formalism\cite{Tancogne-Dejean17}, based on the recently proposed ACBN0 functional~\cite{agapito15}, which can be seen as a pseudo-hybrid reformulation of the density-functional theory plus Hubbard $U$ (DFT+$U$) method.
The time-dependent generalized Kohn-Sham equation within the adiabatic approximation reads (the nonlocal part of the pseudopotential is omitted for conciseness),
\begin{eqnarray}
 i\frac{\partial}{\partial t}|\psi_{n,\mathbf{k}}(t)\rangle = \Big[\frac{(\hat{\mathbf{p}}-\mathbf{A}(t)/c)}{2} + \hat{v}_{\mathrm{ext}} + \hat{v}_{\mathrm{H}}[n(\mathbf{r},t)]\nonumber\\
 + \hat{v}_{\mathrm{xc}}[n(\mathbf{r},t)] + \hat{V}_{U}[n(\mathbf{r},t),\{n^{\sigma\sigma'}_{mm'}\}]\Big]|\psi_{n,\mathbf{k}}(t)\rangle,
\end{eqnarray}
where  $|\psi_{n,\mathbf{k}} \rangle$ is a Pauli spinor representing the Bloch state with a band index $n$, at the point $\mathbf{k}$ in the Brillouin zone, $\hat{v}_{\mathrm{ext}}$ is the ionic potential, $\mathbf{A}(t)$ is the external vector potential describing the laser field,  $\hat{v}_{\mathrm{H}}$ is the Hartree potential, $\hat{v}_{\mathrm{xc}}$ is the exchange-correlation potential, and $\hat{V}^\sigma_{U}$ is the (non-local) operator,
\begin{equation}
\hat{V}_{U}[n,\{n^{\sigma\sigma'}_{mm'}\}] = U_{\mathrm{eff}}\sum_{m,m'}( \frac{1}{2}\delta_{mm'} - n_{mm'} ) \hat{P}_{m,m'}\,.
\label{eq:pot_V_U}
\end{equation}
Here $\hat{P}^{\sigma\sigma'}_{mm'}= |\phi_{m}^\sigma\rangle\langle\phi_{m'}^{\sigma'}|$ is the projector onto the localized subspace defined by the localized orbitals $\{\phi_{m}^\sigma\}$, and $n^{\sigma\sigma'}$ is the density matrix of the localized subspace, both of these quantity been non-diagonal is spin space\cite{Tancogne-Dejean17}. The expressions of $U$ and $J$  can be found for instance in Ref.\cite{Tancogne-Dejean17} for the non-collinear spin case.


\begin{thebibliography}{10}

\bibitem{Kaganov79}
M.~I. Kaganov, I.~M. Lifshitz, Electron theory of metals and geometry.
\newblock {\it Phys. Usp.\/} {\bf 22}, 904-927 (1979).

\bibitem{wu15}
Y.~Wu, N.~H. Jo, M.~Ochi, L.~Huang, D.~Mou, S.~L. Bud'ko, P.~C. Canfield,
  N.~Trivedi, R.~Arita, A.~Kaminski, Temperature-induced Lifshitz transition in
  ${\mathrm{WTe}}_{2}$.
\newblock {\it Phys. Rev. Lett.\/} {\bf 115}, 166602 (2015).

\bibitem{zhang17}
Y.~Zhang, C.~Wang, L.~Yu, G.~Liu, A.~Liang, J.~Huang, S.~Nie, X.~Sun, Y.~Zhang,
  B.~Shen, J.~Liu, H.~Weng, L.~Zhao, G.~Chen, X.~Jia, C.~Hu, Y.~Ding, W.~Zhao,
  Q.~Gao, C.~Li, S.~He, L.~Zhao, F.~Zhang, S.~Zhang, F.~Yang, Z.~Wang, Q.~Peng,
  X.~Dai, Z.~Fang, Z.~Xu, C.~Chen, X.~J. Zhou, Electronic evidence of
  temperature-induced Lifshitz transition and topological nature in $\mathrm{ZrTe_5}$.
\newblock {\it Nature Communications\/} {\bf 8}, 15512 (2017).

\bibitem{chen20}
F.~C. Chen, Y.~Fei, S.~J. Li, Q.~Wang, X.~Luo, J.~Yan, W.~J. Lu, P.~Tong, W.~H.
  Song, X.~B. Zhu, L.~Zhang, H.~B. Zhou, F.~W. Zheng, P.~Zhang, A.~L.
  Lichtenstein, M.~I. Katsnelson, Y.~Yin, N.~Hao, Y.~P. Sun,
  Temperature-induced Lifshitz transition and possible excitonic instability in
  ZrSiSe.
\newblock {\it Phys. Rev. Lett.\/} {\bf 124}, 236601 (2020).

\bibitem{Xiang15}
Z.~J. Xiang, G.~J. Ye, C.~Shang, B.~Lei, N.~Z. Wang, K.~S. Yang, D.~Y. Liu,
  F.~B. Meng, X.~G. Luo, L.~J. Zou, Z.~Sun, Y.~Zhang, X.~H. Chen,
  Pressure-induced electronic transition in black phosphorus.
\newblock {\it Phys. Rev. Lett.\/} {\bf 115}, 186403 (2015).

\bibitem{kang15}
D.~Kang, Y.~Zhou, W.~Yi, C.~Yang, J.~Guo, Y.~Shi, S.~Zhang, Z.~Wang, C.~Zhang,
  S.~Jiang, A.~Li, K.~Yang, Q.~Wu, G.~Zhang, L.~Sun, Z.~Zhao, Superconductivity
  emerging from a suppressed large magnetoresistant state in tungsten
  ditelluride.
\newblock {\it Nature Communications\/} {\bf 6}, 7804 (2015).

\bibitem{sunko19}
V.~Sunko, E.~Abarca~Morales, I.~Markovic, M.~E. Barber, D.~Milosavljevic,
  F.~Mazzola, D.~A. Sokolov, N.~Kikugawa, C.~Cacho, P.~Dudin, H.~Rosner, C.~W.
  Hicks, P.~D.~C. King, A.~P. Mackenzie, Direct observation of a uniaxial
  stress-driven Lifshitz transition in $\mathrm{Sr_2RuO_4}$.
\newblock {\it npj Quantum Materials\/} {\bf 4}, 46 (2019).

\bibitem{burganov16}
B.~Burganov, C.~Adamo, A.~Mulder, M.~Uchida, P.~D.~C. King, J.~W. Harter, D.~E.
  Shai, A.~S. Gibbs, A.~P. Mackenzie, R.~Uecker, M.~Bruetzam, M.~R. Beasley,
  C.~J. Fennie, D.~G. Schlom, K.~M. Shen, Strain control of Fermiology and
  many-body interactions in two-dimensional ruthenates.
\newblock {\it Phys. Rev. Lett.\/} {\bf 116}, 197003 (2016).

\bibitem{Ptok17}
A.~Ptok, K.~J. Kapcia, A.~Cichy, A.~M. Ole{\'{s}}, P.~Piekarz, Magnetic
  Lifshitz transition and its consequences in multi-band iron-based
  superconductors.
\newblock {\it Scientific Reports\/} {\bf 7}, 41979 (2017).

\bibitem{Liu10}
C.~Liu, T.~Kondo, R.~M. Fernandes, A.~D. Palczewski, E.~D. Mun, N.~Ni, A.~N.
  Thaler, A.~Bostwick, E.~Rotenberg, J.~Schmalian, S.~L. Bud'ko, P.~C.
  Canfield, A.~Kaminski, Evidence for a Lifshitz transition in electron-doped
  iron arsenic superconductors at the onset of superconductivity.
\newblock {\it Nature Physics\/} {\bf 6}, 419-423 (2010).

\bibitem{Shi17}
X.~Shi, Z.-Q. Han, X.-L. Peng, P.~Richard, T.~Qian, X.-X. Wu, M.-W. Qiu, S.~C.
  Wang, J.~P. Hu, Y.-J. Sun, H.~Ding, Enhanced superconductivity accompanying a
  Lifshitz transition in electron-doped FeSe monolayer.
\newblock {\it Nature Communications\/} {\bf 8}, 14988 (2017).

\bibitem{Lifshitz60}
I.~M. Lifshitz, Anomalies of electron characteristics of a metal in the high
  pressure region.
\newblock {\it Journal of Experimental and Theoretical Physics\/} {\bf 38},
  1569 (1960).

\bibitem{wang15}
Y.~Wang, M.~N. Gastiasoro, B.~M. Andersen, M.~Tomi\ifmmode~\acute{c}\else
  \'{c}\fi{}, H.~O. Jeschke, R.~Valent\'{\i}, I.~Paul, P.~J. Hirschfeld,
  Effects of Lifshitz transition on charge transport in magnetic phases of
  Fe-based superconductors.
\newblock {\it Phys. Rev. Lett.\/} {\bf 114}, 097003 (2015).

\bibitem{Chi17}
H.~Chi, C.~Zhang, G.~Gu, D.~E. Kharzeev, X.~Dai, Q.~Li, Lifshitz transition
  mediated electronic transport anomaly in bulk $\mathrm{ZrTe_5}$.
\newblock {\it New Journal of Physics\/} {\bf 19}, 015005 (2017).

\bibitem{Xu18}
N.~Xu, Z.~W. Wang, A.~Magrez, P.~Bugnon, H.~Berger, C.~E. Matt, V.~N. Strocov,
  N.~C. Plumb, M.~Radovic, E.~Pomjakushina, K.~Conder, J.~H. Dil, J.~Mesot,
  R.~Yu, H.~Ding, M.~Shi, Evidence of a Coulomb-interaction-induced Lifshitz
  transition and robust hybrid Weyl semimetal in
  ${T}_{d}\text{\ensuremath{-}}{\mathrm{MoTe}}_{2}$.
\newblock {\it Phys. Rev. Lett.\/} {\bf 121}, 136401 (2018).

\bibitem{kutnyakhov20}
D.~Kutnyakhov, R.~P. Xian, M.~Dendzik, M.~Heber, F.~Pressacco, S.~Y. Agustsson,
  L.~Wenthaus, H.~Meyer, S.~Gieschen, G.~Mercurio, A.~Benz, K.~Bühlman,
  S.~Däster, R.~Gort, D.~Curcio, K.~Volckaert, M.~Bianchi, C.~Sanders, J.~A.
  Miwa, S.~Ulstrup, A.~Oelsner, C.~Tusche, Y.-J. Chen, D.~Vasilyev,
  K.~Medjanik, G.~Brenner, S.~Dziarzhytski, H.~Redlin, B.~Manschwetus, S.~Dong,
  J.~Hauer, L.~Rettig, F.~Diekmann, K.~Rossnagel, J.~Demsar, H.-J. Elmers,
  P.~Hofmann, R.~Ernstorfer, G.~Schönhense, Y.~Acremann, W.~Wurth, Time- and
  momentum-resolved photoemission studies using time-of-flight momentum
  microscopy at a free-electron laser.
\newblock {\it Review of Scientific Instruments\/} {\bf 91}, 013109 (2020).

\bibitem{Tancogne-Dejean17}
N.~Tancogne-Dejean, M.~J.~T. Oliveira, A.~Rubio, Self-consistent
  $\mathrm{DFT}+U$ method for real-space time-dependent density functional
  theory calculations.
\newblock {\it Phys. Rev. B\/} {\bf 96}, 245133 (2017).

\bibitem{Tancogne-Dejean18}
N.~Tancogne-Dejean, M.~A. Sentef, A.~Rubio, Ultrafast modification of Hubbard
  $U$ in a strongly correlated material: Ab initio high-harmonic generation in
  NiO.
\newblock {\it Phys. Rev. Lett.\/} {\bf 121}, 097402 (2018).

\bibitem{Topp18}
G.~E. Topp, N.~Tancogne-Dejean, A.~F. Kemper, A.~Rubio, M.~A. Sentef,
  All-optical nonequilibrium pathway to stabilising magnetic Weyl semimetals in
  pyrochlore iridates.
\newblock {\it Nature Communications\/} {\bf 9}, 4452 (2018).

\bibitem{soluyanov15}
A.~A. Soluyanov, D.~Gresch, Z.~Wang, Q.~Wu, M.~Troyer, X.~Dai, B.~A. Bernevig,
  Type-II Weyl semimetals.
\newblock {\it Nature\/} {\bf 527}, 495-498 (2015).

\bibitem{wan11}
X.~Wan, A.~M. Turner, A.~Vishwanath, S.~Y. Savrasov, Topological semimetal and
  Fermi-arc surface states in the electronic structure of pyrochlore iridates.
\newblock {\it Phys. Rev. B\/} {\bf 83}, 205101 (2011).

\bibitem{xu15}
S.-Y. Xu, I.~Belopolski, N.~Alidoust, M.~Neupane, G.~Bian, C.~Zhang, R.~Sankar,
  G.~Chang, Z.~Yuan, C.-C. Lee, S.-M. Huang, H.~Zheng, J.~Ma, D.~S. Sanchez,
  B.~Wang, A.~Bansil, F.~Chou, P.~P. Shibayev, H.~Lin, S.~Jia, M.~Z. Hasan,
  Discovery of a Weyl fermion semimetal and topological Fermi arcs.
\newblock {\it Science\/} {\bf 349}, 613--617 (2015).

\bibitem{hosur13}
P.~Hosur, X.~Qi, Recent developments in transport phenomena in Weyl semimetals.
\newblock {\it Comptes Rendus Physique\/} {\bf 14}, 857 - 870 (2013).

\bibitem{Aryal19}
N.~Aryal, E.~Manousakis, Importance of electron correlations in understanding
  photoelectron spectroscopy and Weyl character of ${\mathrm{MoTe}}_{2}$.
\newblock {\it Phys. Rev. B\/} {\bf 99}, 035123 (2019).

\bibitem{Puppin19}
M.~Puppin, Y.~Deng, C.~W. Nicholson, J.~Feldl, N.~B.~M. Schröter, H.~Vita,
  P.~S. Kirchmann, C.~Monney, L.~Rettig, M.~Wolf, R.~Ernstorfer, Time- and
  angle-resolved photoemission spectroscopy of solids in the extreme
  ultraviolet at 500 kHz repetition rate.
\newblock {\it Review of Scientific Instruments\/} {\bf 90}, 023104 (2019).

\bibitem{Medjanik17}
K.~Medjanik, O.~Fedchenko, S.~Chernov, D.~Kutnyakhov, M.~Ellguth, A.~Oelsner,
  B.~Sch{\"o}nhense, T.~R.~F. Peixoto, P.~Lutz, C.-H. Min, F.~Reinert,
  S.~D{\"a}ster, Y.~Acremann, J.~Viefhaus, W.~Wurth, H.~J. Elmers,
  G.~Sch{\"o}nhense, Direct 3D mapping of the Fermi surface and Fermi velocity.
\newblock {\it Nature Materials\/} {\bf 16}, 615-621 (2017).

\bibitem{Maklar20}
J.~Maklar, S.~Dong, S.~Beaulieu, T.~Pincelli, M.~Dendzik, Y.~W. Windsor, R.~P.
  Xian, M.~Wolf, R.~Ernstorfer, L.~Rettig, A quantitative comparison of
  time-of-flight momentum microscopes and hemispherical analyzers for
  time-resolved ARPES experiments.
  \newblock {\it arXiv:2008.05829 [physics.ins-det]\/} (2020).

\bibitem{zhang19}
M.~Y. Zhang, Z.~X. Wang, Y.~N. Li, L.~Y. Shi, D.~Wu, T.~Lin, S.~J. Zhang, Y.~Q.
  Liu, Q.~M. Liu, J.~Wang, T.~Dong, N.~L. Wang, Light-induced subpicosecond
  lattice symmetry switch in ${\mathrm{MoTe}}_{2}$.
\newblock {\it Phys. Rev. X\/} {\bf 9}, 021036 (2019).

\bibitem{kiwia1975low}
H.~L. Kiwia, E.~F. Westrum~Jr, Low-temperature heat capacities of molybdenum
  diselenide and ditelluride.
\newblock {\it The Journal of Chemical Thermodynamics\/} {\bf 7}, 683--691
  (1975).

\bibitem{chen16}
F.~C. Chen, H.~Y. Lv, X.~Luo, W.~J. Lu, Q.~L. Pei, G.~T. Lin, Y.~Y. Han, X.~B.
  Zhu, W.~H. Song, Y.~P. Sun, Extremely large magnetoresistance in the type-II
  Weyl semimetal $\mathrm{Mo}{\mathrm{Te}}_{2}$.
\newblock {\it Phys. Rev. B\/} {\bf 94}, 235154 (2016).

\bibitem{yankowitz19}
M.~Yankowitz, S.~Chen, H.~Polshyn, Y.~Zhang, K.~Watanabe, T.~Taniguchi,
  D.~Graf, A.~F. Young, C.~R. Dean, Tuning superconductivity in twisted bilayer
  graphene.
\newblock {\it Science\/} {\bf 363}, 1059--1064 (2019).

\bibitem{xian19}
R.~P. Xian, Y.~Acremann, S.~Y. Agustsson, M.~Dendzik, K.~Bühlmann, D.~Curcio,
  D.~Kutnyakhov, F.~Pressacco, M.~Heber, S.~Dong, J.~Demsar, W.~Wurth,
  P.~Hofmann, M.~Wolf, L.~Rettig, R.~Ernstorfer, An open-source, distributed
  workflow for band mapping data in multidimensional photoemission spectroscopy.
  \newblock {\it arXiv:1909.07714 [physics.data-an]\/} (2019).

\bibitem{Qi16}
Y.~Qi, P.~G. Naumov, M.~N. Ali, C.~R. Rajamathi, W.~Schnelle, O.~Barkalov,
  M.~Hanfland, S.-C. Wu, C.~Shekhar, Y.~Sun, V.~S{\"u}{\ss}, M.~Schmidt,
  U.~Schwarz, E.~Pippel, P.~Werner, R.~Hillebrand, T.~F{\"o}rster, E.~Kampert,
  S.~Parkin, R.~J. Cava, C.~Felser, B.~Yan, S.~A. Medvedev, Superconductivity
  in Weyl semimetal candidate $\mathrm{MoTe_2}$.
\newblock {\it Nature Communications\/} {\bf 7}, 11038 (2016).

\bibitem{1912.07921}
N.~Tancogne-Dejean, M.~J.~T. Oliveira, X.~Andrade, H.~Appel, C.~H. Borca, G.~L.
  Breton, F.~Buchholz, A.~Castro, S.~Corni, A.~A. Correa, U.~D. Giovannini,
  A.~Delgado, F.~G. Eich, J.~Flick, G.~Gil, A.~Gomez, N.~Helbig, H.~Hübener,
  R.~Jestädt, J.~Jornet-Somoza, A.~H. Larsen, I.~V. Lebedeva, M.~Lüders,
  M.~A.~L. Marques, S.~T. Ohlmann, S.~Pipolo, M.~Rampp, C.~A. Rozzi, D.~A.
  Strubbe, S.~A. Sato, C.~Schäfer, I.~Theophilou, A.~Welden, A.~Rubio,
  Octopus, a computational framework for exploring light-driven phenomena and
  quantum dynamics in extended and finite systems.
  \newblock {\it J. Chem. Phys.\/} {\bf 152}, 124119 (2020).

\bibitem{Andrade15}
X.~Andrade, D.~Strubbe, U.~De~Giovannini, A.~H. Larsen, M.~J.~T. Oliveira,
  J.~Alberdi-Rodriguez, A.~Varas, I.~Theophilou, N.~Helbig, M.~J. Verstraete,
  L.~Stella, F.~Nogueira, A.~Aspuru-Guzik, A.~Castro, M.~A.~L. Marques,
  A.~Rubio, Real-space grids and the octopus code as tools for the development
  of new simulation approaches for electronic systems.
\newblock {\it Phys. Chem. Chem. Phys.\/} {\bf 17}, 31371-31396 (2015).

\bibitem{castro06}
A.~Castro, H.~Appel, M.~Oliveira, C.~A. Rozzi, X.~Andrade, F.~Lorenzen,
  M.~A.~L. Marques, E.~K.~U. Gross, A.~Rubio, Octopus: a tool for the
  application of time-dependent density functional theory.
\newblock {\it physica status solidi (b)\/} {\bf 243}, 2465-2488 (2006).

\bibitem{agapito15}
L.~A. Agapito, S.~Curtarolo, M.~Buongiorno~Nardelli, Reformulation of
  $\mathrm{DFT}+U$ as a pseudohybrid hubbard density functional for accelerated
  materials discovery.
\newblock {\it Phys. Rev. X\/} {\bf 5}, 011006 (2015).

\bibitem{crepaldi17}
A.~Crepaldi, G.~Aut\`es, G.~Gatti, S.~Roth, A.~Sterzi, G.~Manzoni,
  M.~Zacchigna, C.~Cacho, R.~T. Chapman, E.~Springate, E.~A. Seddon, P.~Bugnon,
  A.~Magrez, H.~Berger, I.~Vobornik, M.~Kall\"ane, A.~Quer, K.~Rossnagel,
  F.~Parmigiani, O.~V. Yazyev, M.~Grioni, Enhanced ultrafast relaxation rate in
  the Weyl semimetal phase of $\mathrm{MoTe}_2$ measured by time- and
  angle-resolved photoelectron spectroscopy.
\newblock {\it Phys. Rev. B\/} {\bf 96}, 241408 (2017).

\bibitem{pizzi20}
G.~Pizzi, V.~Vitale, R.~Arita, S.~Blügel, F.~Freimuth, G.~G{\'{e}}ranton,
  M.~Gibertini, D.~Gresch, C.~Johnson, T.~Koretsune, J.~Iba{\~{n}}ez-Azpiroz,
  H.~Lee, J.-M. Lihm, D.~Marchand, A.~Marrazzo, Y.~Mokrousov, J.~I. Mustafa,
  Y.~Nohara, Y.~Nomura, L.~Paulatto, S.~Ponc{\'{e}}, T.~Ponweiser, J.~Qiao,
  F.~Thöle, S.~S. Tsirkin, M.~Wierzbowska, N.~Marzari, D.~Vanderbilt,
  I.~Souza, A.~A. Mostofi, J.~R. Yates, Wannier90 as a community code: new
  features and applications.
\newblock {\it Journal of Physics: Condensed Matter\/} {\bf 32}, 165902 (2020).

\bibitem{de2016first}
U.~De~Giovannini, H.~Hubener, A.~Rubio, A first-principles time-dependent
  density functional theory framework for spin and time-resolved
  angular-resolved photoelectron spectroscopy in periodic systems.
\newblock {\it Journal of chemical theory and computation\/} {\bf 13}, 265--273
  (2016).

\end{thebibliography}

\noindent \textbf{Funding:} This work was funded by the Max Planck Society, the European Research Council (ERC) under the European Union’s Horizon 2020 research and innovation program (Grant No. ERC-2015-CoG-682843, ERC-2015-AdG694097 and H2020-FETOPEN-2018-2019-2020-01 (OPTOLogic - grant agreement No. 899794)), the Grupos Consolidados (IT1249-19) and the Deutsche Forschungsgemein-schaft (DFG, German Research Foundation) within the Emmy Noether program (Grant No. RE 3977/1 and MS 2558/2-1), the Cluster of Excellence "Advanced Imaging of Matter" (AIM), the SFB925 "Light induced dynamics and control of correlated quantum systems", the Collaborative Research Center/Transregio 227 "Ultrafast Spin Dynamics" (project B07 and A09), the FOR1700 project (Project E5) and the Priority Program SPP 2244 (project No.~443366970).  The Flatiron Institute is a division of the Simons Foundation. S.B. acknowledges financial support from the NSERC-Banting Postdoctoral Fellowships Program.\\

\noindent \textbf{Author Contributions} S.B., S.D., M.D., J.M., T.P. and L.R. performed the time-resolved multidimensional photoemission spectroscopy experiments. S.B. analyzed the experimental data. R.P.X developed the open source workflow for data preprocessing. R.E., L.R. and M.W. were responsible for developing the infrastructures allowing these measurements as well as for the overall project direction. N.T.D. performed the theoretical calculations, their analysis and interpretation, with the guidance of M.A.S. and A.R. S.B. wrote the first draft with inputs from N.T.D., M.A.S., and R.E. All authors contributed to the discussions and the final version of the manuscript.\\

\noindent \textbf{Competing Interests} The authors declare that they have no competing interests.\\
\noindent \textbf{Data and materials availability:} All data needed to evaluate the conclusions in the paper are present in the paper and/or the Supplementary Materials.

\clearpage

\begin{centering}
\huge{\textbf{Supplementary Materials}}
\end{centering}

\begin{center}
\textbf{I. Additional experimental data}
\end{center}

\textbf{Dynamics along X-$\mathrm{\Gamma}$-X}

\begin{figure}[h!]
\centering\includegraphics[width=0.8\textwidth]{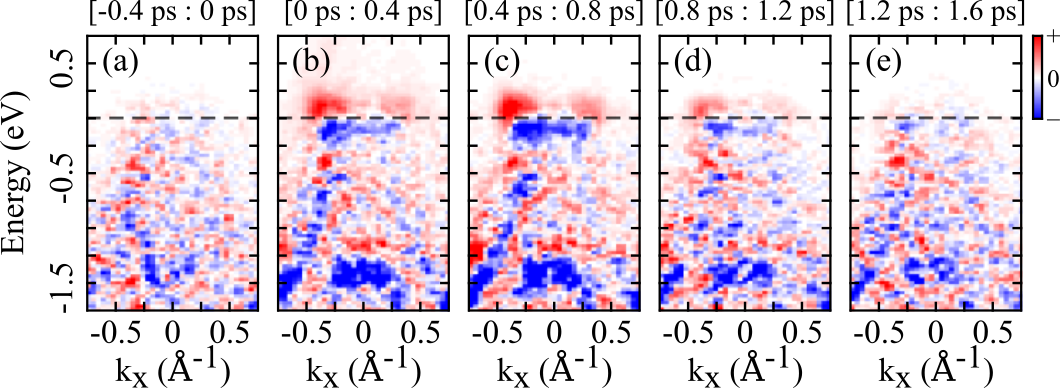}
\caption{\textbf{Ultrafast electronic dynamic along X-$\Gamma$-X.} (a)-(e) Differential (signal before time-zero subtracted) energy-resolved cuts along X-$\mathrm{\Gamma}$-X ($\mathrm{k_y=0}$) as a function of time delay between the IR pump and the XUV probe. These data are extracted from the same scan as for Fig.~2 of the main text. The pump pulse intensity is estimated to be 6.7x10$^9$ $\mathrm{W/cm^2}$.}
\label{SM_Diff_kx}
\end{figure}

In Fig.~2 of the main paper, we presented differential Fermi surfaces and cuts along Y-$\mathrm{\Gamma}$-Y ($\mathrm{k_x=0}$) direction, to show the unambiguous signature of Lifshitz transition. Our multidimensional detection scheme allows us to measure differential signals along any direction of the Brillouin zone with sufficient counting statistics. To compare our data with previously published time- and angle-resolved photoemission spectroscopy (ARPES) results on the same material, we can plot the photoemission differential signal along X-$\mathrm{\Gamma}$-X ($\mathrm{k_y=0}$) high-symmetry direction, as reported by Crepaldi \textit{et al.} \cite{crepaldi17}. 
The strong enhancement of the signal around $\pm$ 0.25 $\mathrm{\AA^{-1}}$ above the Fermi level, the depletion around $\mathrm{\Gamma}$ below the Fermi level, as well as the extracted excited states lifetime of $\sim$ 270 fs, from our experimental data presented on Fig. \ref{SM_Diff_kx}, is in good agreement with the observations of Crepaldi \textit{et al.} \cite{crepaldi17}, who studied the enhanced ultrafast relaxation rate in the Weyl semimetallic (low temperature) phase of $\mathrm{MoTe_2}$. In that paper, the authors didn't report the dynamics along $\mathrm{\Gamma}$-Y, which is the direction along which the signature of the Lifshitz transition is the most evident. 

\textbf{Time-dependent electronic temperature}

We have extracted the time-dependent electronic temperature, to estimate the heating of the electronic system by the pump pulse. This information provides a reference point for theoretical simulations, presented in the next sections. To do so, we fitted the energy distribution-curve (EDCs) (integrated along $\mathrm{\Gamma}$-X) measured experimentally by Fermi-Dirac distribution functions convoluted with a Gaussian function as the spectral instrument response function. We have used EDCs obtained by integrating the signal along $\mathrm{\Gamma}$-X, in order to exclude the effect of the time-dependent density of states along $\mathrm{\Gamma}$-Y, which could introduce some ambiguity in the fitting procedure. 

\begin{figure}[t!]
\centering\includegraphics[width=0.9\textwidth]{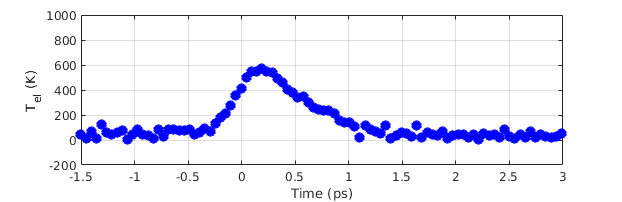}
\caption{\textbf{Time-dependent electronic temperature.}: Electronic temperature as a function of pump-probe delay, extracted from fitting the energy distribution-curve (integrated along $\mathrm{\Gamma}$-X) by a resolution-broadened Fermi-Dirac distribution function.}
\label{TDElectronicTemperature}
\end{figure}

First, we have fixed the electronic temperature of the Fermi-Dirac distribution to 30 K and kept the Gaussian width as a free parameter of the fit, for the case of the unpumped system, to estimate the instrument response function. Next, we have fixed the instrument response function (the Gaussian width), and fit a Fermi-Dirac distribution for each pump-probe delay, keeping the electronic temperature as a free parameter of the fit. The results of the fit are shown in Fig.~\ref{TDElectronicTemperature}.

As shown in Fig.~\ref{TDElectronicTemperature}, after the interaction with the pump laser pulse, the electronic temperature increases to almost 600 K and then gradually recovers to the original temperature of $\sim$ 30 K, within a time-scale similar to the excited states population lifetime (signal above the Fermi level). As explained in the theoretical section of the Supplementary Information, this increase of the electronic temperature (dynamical populations), combined with the dynamical modification of Hubbard $U_{\mathrm{eff}}$, is at the origin of the ultrafast non-equilibrium Lifshitz transition.  

\textbf{Fitting the $\gamma$-pocket position}

\begin{figure}[t!]
\centering\includegraphics[width=0.4\textwidth]{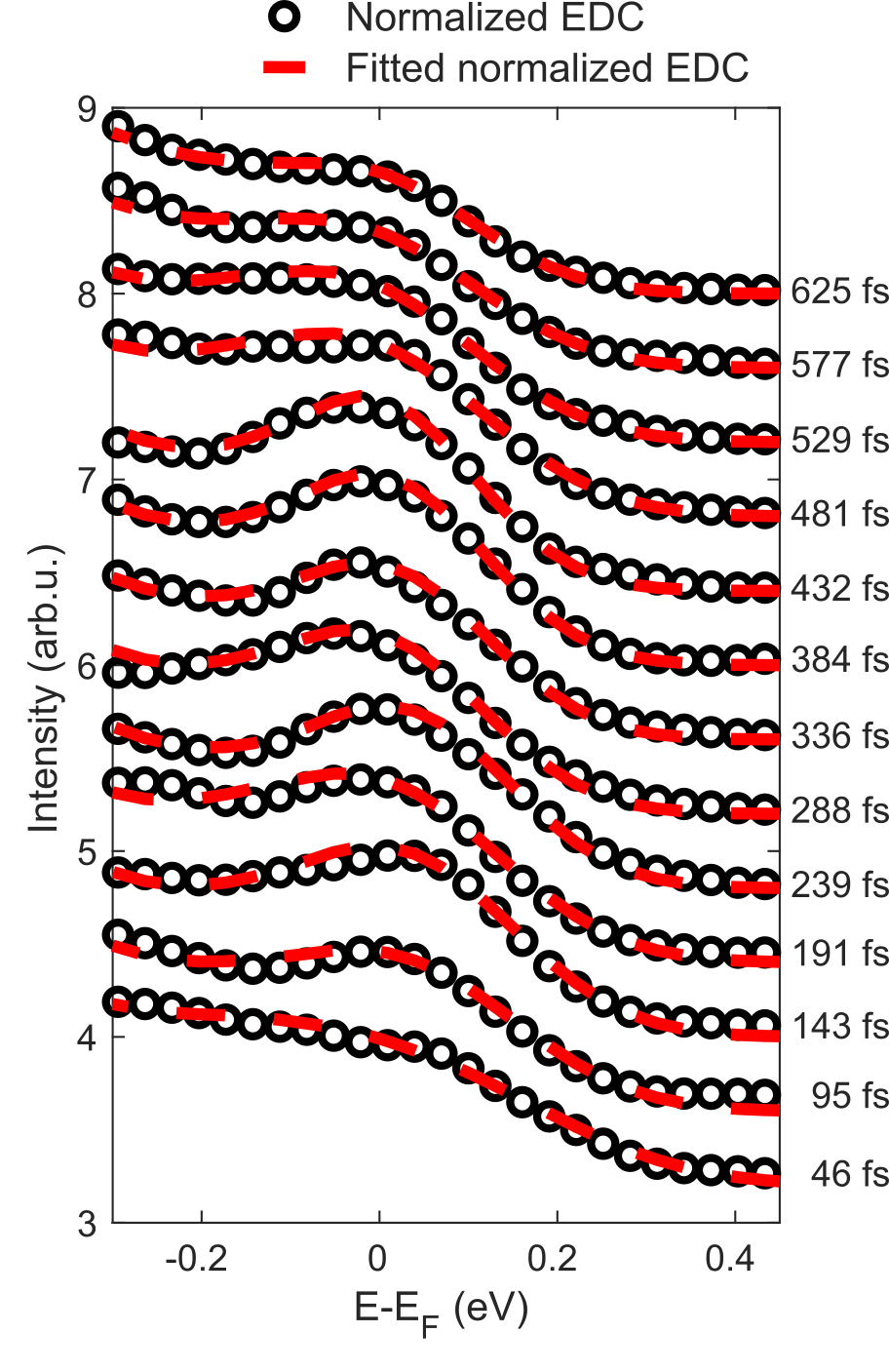}
\caption{\textbf{Energy distribution curves fitting}: The black dots are normalized experimentally measured EDC cuts at the Y point of the Brillouin zone, for different pump-probe delays (indicated on the right side of the figure). The dashed red lines are the results of the EDC fitting.}
\label{EDC}
\end{figure}

In order to extract the position of the bottom of the $\gamma$-pocket for different pump-probe delays, we have fitted the energy distribution curves, taken at the Y point of the Brillouin zone. We have fitted two Gaussian separated by 36.5 meV (upper and lower bands at Y) weighted by the instrument response function broaden time-dependent Fermi-Dirac distribution extracted from the fit of the EDC along $\mathrm{\Gamma}$-X (region where the band structure is not modified upon photoexcitation). Examples of such EDCs fitting are shown in Fig. \ref{EDC}.

\newpage

\begin{center}
\textbf{II. Adiabatic Lifshitz transition}
\end{center}

We first theoretically analyze the adiabatic Lifshitz transition and in particular the effect of $U_{\mathrm{eff}}$ on the energy position of the $\gamma$-pocket. For this, we performed a set of DFT+$U$ calculations while varying $U_{\mathrm{eff}}$. Our results are shown in Fig.~1c in the main text. The self-consistently evaluated $U_{\mathrm{eff}}$ is found to be 2.05\,eV. From our calculations, we find that the Lifshitz transition occurs around 1.5\,eV, which is consistent with the value obtained from a similar analysis by Xu \textit{et al.}~\cite{Xu18}.

From these data, we find that in the adiabatic limit, for a ~70\,meV change in position of the $\gamma$-pocket, which has been measured experimentally, (see Fig.~3c in the main text), a change in $U_{\mathrm{eff}}$ of 1.8\,eV is required. This is in clear contrast with the results obtained in Fig.~3e from our TDDFT+$U$ calculations, from which the change in $U_{\mathrm{eff}}$ at the experimental intensity is estimated to be around 30\,meV only. Thus, our results indicate that the measured energy downshifts of the pocket, ultimately leading to the Lifshitz transition, cannot originate only from an adiabatic change of $U_{\mathrm{eff}}$ and that more subtle dynamics are at play here. 

Note that we carefully checked that the laser-induced changes in $U_{\mathrm{eff}}$ are converged and that adding more $\mathbf{k}$-point to sample the Brillouin zone do not lead to sizable differences in the dynamics of $U_{\mathrm{eff}}$. Moreover, note that due to numerical artifacts, the Fig.~3a and Fig.~3b are obtained without spin-orbit coupling included. Whereas the main results are not affected by the inclusion or not of the spin-orbit coupling, the Wannierization procedure, employed to get the Fermi surface cut from the non-equilibrium states, is found to be less stable when the spin-orbit coupling is included.

We want to point out the analogy to the creation of a Weyl semimetallic phase induced by a laser in pyrochlore iridates, that some of us predicted \cite{Topp18}. Indeed, in this previous work, we showed that light-induced reduction of $U_{\mathrm{eff}}$ is leading to the transient appearance of a Weyl semimetallic phase in pyrochlore irridate for values of $U_{\mathrm{eff}}$ for which, according to the equilibrium (adiabatic) phase diagram, should not allow it. 

The experimentally measured change in the pocket position also shows a clear delay between the change in the pocket energy position and the laser pulse itself (timed by the LAPE signal, Fig.~3b in the main text), indicating that some transient dynamics still takes place after the end of the laser pulse and before the material starts to returns to its ground state due to thermalization effects. This is an indication that the system is brought to a non-equilibrium state at the end of the laser pulse.

As discussed below, our results provide evidence explaining the observed light-induced Lifshitz transition, but we found that one needs to take into account the non-equilibrium dynamics induced by the laser and not only a simpler adiabatic change of the Hubbard $U_{\mathrm{eff}}$.

\begin{center}
\textbf{III. Non-equilibrium Fermi surface}
\end{center}

From the theoretical point of view, the Fermi surface is a concept that is well defined at equilibrium, and for a vanishing temperature. However, in time- and angle-resolved photoemission experiments, it is possible to measure the Fermi surface evolving in time for a fixed Fermi energy, taking as reference the equilibrium value. Here, we aim at comparing this measured non-equilibrium Fermi surface to the results of our time-dependent simulations, which include both the change in populations and time-evolved Hubbard $U$. We employed for this the non-equilibrium states taken after the end of the laser pulse and computed the corresponding energies
\begin{equation}
    E^{\mathrm{neq}}_{n\mathbf{k}}(t) = \langle \psi_{n\mathbf{k}}(t) | \hat{H}(t) | \psi_{n\mathbf{k}}(t) \rangle,
\end{equation}
where $n$ refers to a band index, $\mathbf{k}$ is the $\mathbf{k}$-point index, $\psi_{n\mathbf{k}}(t)$ the Pauli spinor representing the Bloch state obtained from the time-evolution of the TDDFT Kohn-Sham equations, and $\hat{H}[n(t)](t)$ is the Hamiltonian constructed from the time-evolved electronic density $n(t)$ as well as $U_{\mathrm{eff}}(t)$ and the corresponding non-equilibrium occupations $n_{mm'}^{\sigma\sigma'}(t)$.

We note that using these energies after the end of the laser pulse (\textit{i.e.} when there is no vector potential due to the laser) makes the result of our analysis gauge invariant. Alternatively, one might want to use the adiabatic eigenvalues, $E^{ad}_{n\mathbf{k}}(t)$, of the time-evolved Hamiltonian, $\hat{H}(t)$, to investigate the non-equilibrium Fermi surface defined by
\begin{equation}
   \hat{H}(t) | \psi^{\mathrm{ad}}_{n\mathbf{k}}\rangle =  E^{\mathrm{ad}}_{n\mathbf{k}}(t) | \psi^{\mathrm{ad}}_{n\mathbf{k}}\rangle.
\end{equation}
Here, similar to the previous case, $\hat{H}(t)$ is the Hamiltonian constructed from the time-evolved electronic density $n(t)$ as well as $U_{\mathrm{eff}}(t)$ and the corresponding non-equilibrium occupations $n_{mm'}^{\sigma\sigma'}(t)$.

We found that the adiabatic states do not show any Lifshitz transition compared to the non-equilibrium states, which is fully consistent with our analysis above. Indeed, if the adiabatic states would show a Lifshitz transition, the band structure corresponding to the Hamiltonian constructed from $U(t)$ would show a Lifshitz transition, which we have shown not to be the case for the excitation density used in this experiment. This is a strong indication of the non-adiabatic nature of the measured Lifshitz transition.

\begin{center}
\textbf{IV. Effect of dynamical $U$}
\end{center}

We further analyze the role of the different ingredients to reach the Lifshitz transition. From the theoretical results presented in the main text, the change of $U_{\mathrm{eff}}$ is found to be quite small in Fig.~3e, compared to what could be expected from the equilibrium phase diagram (Fig.~1c in the main text), in order to reach the Lifshitz transition. To disentangle the different effects taking place, we also performed a time-dependent simulation with a frozen $U_{\mathrm{eff}}$ and compared with the result with a time-evolving $U_{\mathrm{eff}}$ presented in the main text. This analysis allows us to clearly identify the role of dynamical $U_{\mathrm{eff}}$ in driving the Lifshitz transition. Indeed, for a time-dependent $U_{\mathrm{eff}}$-frozen simulation, the non-adiabatic nature of the time-evolved states is still taken into account, as well as dynamical populations, as described by the local-density approximation, but the dynamical correlations (captured by time-evolving $U_{\mathrm{eff}}$) are not included.

For a clearer comparison with the experimental results, we simulate a differential angle-resolved photoemission signal from the excited-state eigenvalues. We employed Wannier90 \cite{pizzi20} to interpolate the band structures on a finer $\mathbf{k}$-point grid. We, then, convoluted in energy and in momentum by a Gaussian whose widths correspond to the experimental resolutions, respectively 120\,meV and 0.05$\mathrm{\AA^{-1}}$.

\begin{figure}[t!]
\centering\includegraphics[width=0.8\textwidth]{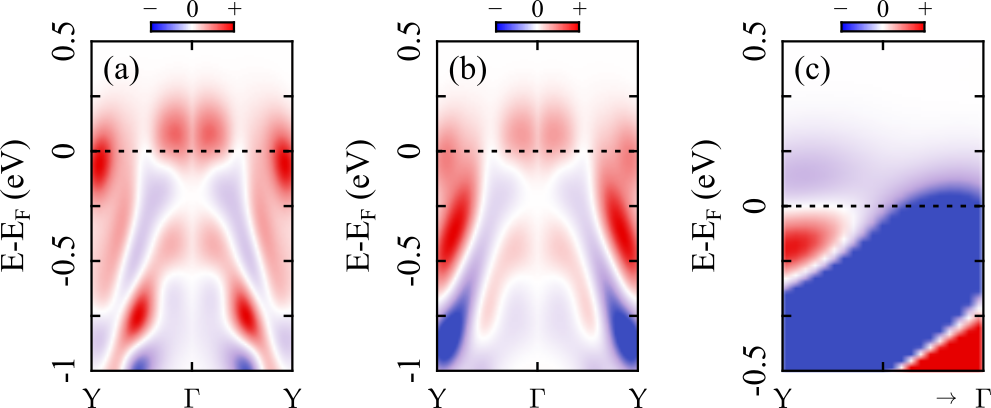}
\caption{\textbf{Effect of dynamical $U_{\mathrm{eff}}$ along Y-$\Gamma$-Y.} a) Calculation performed from the excited states obtained at the end of the laser pulse for a dynamical $U_{\mathrm{eff}}$ calculation. b) Same as a) but for a frozen $U_{\mathrm{eff}}$ calculation.  The (a) and (b) panels correspond to simulated differential maps, similar to what is measured experimentally. c) Difference of the two signals, \textit{i.e.} (a) and (b), close to the position of the pocket.}
\label{fig:ARPES}
\end{figure}

Our results are shown in Fig.~\ref{fig:ARPES} for the Y-$\Gamma$-Y direction, \textit{i.e.} the direction where the hallmark of the Lifshitz transition, and we found that $\gamma$-pocket crossing the Fermi level, appears. From these data, it is clear that the appearance of the $\gamma$-pocket in the angle-resolved photoemission signal crucially depends on the dynamical $U_{\mathrm{eff}}$. Indeed, in Fig.~\ref{fig:ARPES}(b), the signal of the pocket does not appear clearly, whereas it is the dominating feature in Fig.~\ref{fig:ARPES}(a).  As evidenced by the difference close to the position of the $\gamma$-pocket, which displays a clear differential profile, the dynamical $U_{\mathrm{eff}}$ calculation yields a pocket below the Fermi energy whereas the frozen $U_{\mathrm{eff}}$ approach yields a pocket at higher energy. In other words, as shown in Fig.~\ref{fig:ARPES}(c), the dynamical $U_{\mathrm{eff}}$ case leads to such a down-shift compared to the frozen U calculation, and the differential profile is occurring at the Fermi energy. This clearly shows that the dynamical $U_{\mathrm{eff}}$ leads to a down-shift of the pocket below the Fermi energy, whereas the frozen $U_{\mathrm{eff}}$ does not. Otherwise, the differential profile would be centered around a value below the Fermi energy.

This is a clear indication that the dynamical correlations are a crucial ingredient to observe the Lifshitz transition.

\begin{figure}[t!]
\centering\includegraphics[width=0.55\textwidth]{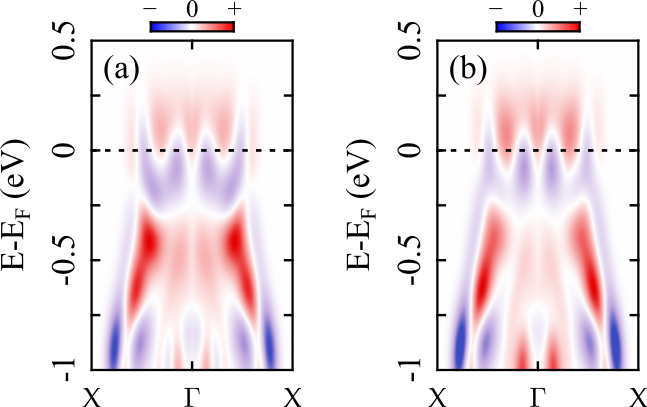}
\caption{\textbf{Effect of dynamical $U_{\mathrm{eff}}$ along X-$\Gamma$-X.} a) Calculation performed from the excited states obtained at the end of the laser pulse for a dynamical $U_{\mathrm{eff}}$ calculation. b) Same as a) but for a frozen $U_{\mathrm{eff}}$ calculation. These panels correspond to simulated differential maps, similar to what is measured experimentally.}
\label{fig:ARPES_GX}
\end{figure}

We also computed similar plots for the X-$\Gamma$-X direction (see Fig.~\ref{fig:ARPES_GX}) for which $U_{\mathrm{eff}}$ is expected not to play an important role. As expected, we do not observe any qualitative difference between the frozen and dynamical $U_{\mathrm{eff}}$ calculations, confirming the validity of our approach. 

Importantly, in both cases, we note that the simulated ARPES spectra are qualitative agreement with the experimental ARPES signals (see Fig.~2 in the main text and Fig.~\ref{SM_Diff_kx}). Finally, we note that these results employ here a crude approximation to the measured ARPES signal, in which transition dipole matrix elements are not taken into account, which would correspond to more intricate and numerically more expensive TDDFT calculations in supercell~\cite{de2016first}.

\begin{center}
\textbf{V. Note on the role of dynamical populations}
\end{center}

As shown in the main text, the full TDDFT+$U_{\mathrm{eff}}$ simulations, which include both the dynamical evolution of $U_{\mathrm{eff}}$ and of the populations, predict the ultrafast Lifshitz transition. Moreover, the comparison between the non-equilibrium simulations performed with frozen and with dynamical $U_{\mathrm{eff}}$ (Fig.~\ref{fig:ARPES}) showed that even if Hubbard $U_{\mathrm{eff}}$ do not reach the adiabatic critical value ($U_{\mathrm{eff}}$ $\sim$ 1.5 eV) to reach the Lifshitz transition, the inclusion of dynamical $U_{\mathrm{eff}}$ is essential to reach the non-equilibrium Lifshitz transition. By taking into account these observations, we argue that both the dynamical modification of the $U_{\mathrm{eff}}$, and of the populations, are necessary to reach the ultrafast non-equilibrium Lifshitz transition. This also implies that the novel non-equilibrium route requires a significantly smaller change in the Hubbard $U_{\mathrm{eff}}$ than the adiabatic case to reach the Lifshitz transition.

\begin{center}
\textbf{VI. Band structure of $\mathrm{1T'}$-$\mathrm{MoTe_2}$}
\end{center}

\begin{figure}[t!]
\centering\includegraphics[width=0.5\textwidth]{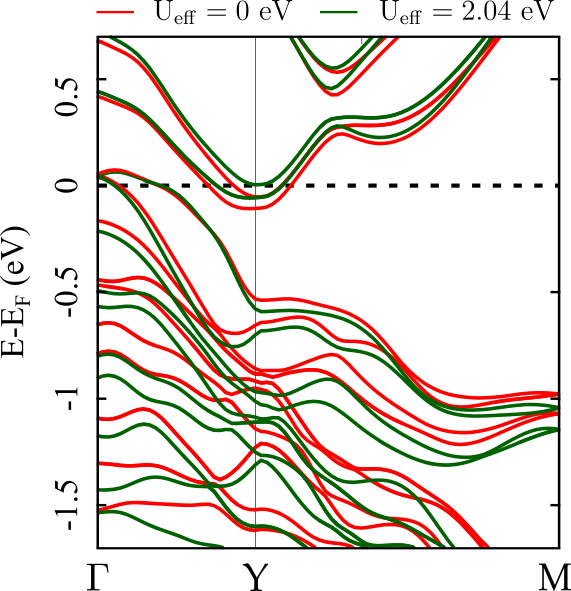}
\caption{\textbf{Band structure of \textbf{$\mathrm{1T'}$-$\mathrm{MoTe_2}$}}. The electronic band structure of $\mathrm{1T'}$-$\mathrm{MoTe_2}$ for the equilibrium self-consistent value of $U_{\mathrm{eff}}$ = 2.04 eV (in green) and for the reduced value of $U_{\mathrm{eff}}$ = 0 eV.}
\label{fig:ARPES_1T}
\end{figure}

We also performed simulations for the $\mathrm{1T'}$-phase of MoTe$_2$ (see Fig.~\ref{fig:ARPES_1T}), which is the high temperature phase. We employed the same parameters as for equilibrium studies of T$_{\mathrm{d}}$-MoTe$_2$, and used the previously reported atomic coordinates~\cite{Qi16}.
We performed both DFT and DFT+$U$ calculations, in which we evaluated the \textit{ab initio} $U_{\mathrm{eff}}$, see Sec. III for more details. As an important result, we found that even including a Hubbard $U_{\mathrm{eff}}$ of 2.04\,eV (obtained from first principles), the $\gamma$-pocket of the $\mathrm{1T'}$ phase of MoTe$_2$ is located below the Fermi energy. 

This shows that at equilibrium, the $\mathrm{1T'}$- and the T$_{\mathrm{d}}$-phase have different Fermi surface topologies. This raises the question of whether the measured data corresponds to a light-induced structural phase transition to the high-temperature $\mathrm{1T'}$ phase, as observed in Ref.~\cite{zhang19}. We argue in detail in the main text that this is not the case. Briefly, the pump fluence that we used ($\sim$ 0.6 mJ/$\mathrm{cm^2}$) is significantly lower than the structural phase transition critical fluence ($>$2 mJ/$\mathrm{cm^2}$), and the time-scales relevant for the experimentally observed Lifshitz transition are completely different than the ones associated with the light-induced structural phase transition \cite{zhang19}.

\clearpage

\end{document}